%% file: SBLS.tex
\documentclass[twocolumn]{aastex63}

\usepackage{float}
\usepackage{amsmath}
\usepackage[boxed,vlined,figure]{algorithm2e}
\usepackage[noend]{algpseudocode}
\usepackage{graphicx}
\usepackage{xcolor}
\usepackage{nicefrac}
\usepackage{txfonts}

\renewcommand{\arraystretch}{1.2}

\defcitealias{Kovacs2002}{KZM}

\begin{document}

\title{Sparse Box-fitting Least Squares}

\author[0000-0001-5850-4373]{Aviad Panahi}
\affiliation{School of Physics and Astronomy, Raymond and Beverly Sackler Faculty of Exact Sciences, Tel Aviv University, Tel Aviv 6997801, Israel} 
\author[0000-0003-3173-3138]{Shay Zucker}
\affiliation{Porter School of the Environment and Earth Sciences, Raymond and Beverly Sackler Faculty of Exact Sciences, Tel Aviv University, Tel Aviv 6997801, Israel}
\begin{abstract}
We present a new implementation of the commonly used Box-fitting Least Squares (BLS) algorithm, for the detection of transiting exoplanets in photometric data. Unlike BLS, our new implementation -- Sparse BLS (SBLS), does not use binning of the data into phase bins, nor does it use any kind of phase grid. Thus, its detection efficiency does not depend on the transit phase, and is therefore slightly better than that of BLS. For sparse data, it is also significantly faster than BLS. It is therefore perfectly suitable for large photometric surveys producing unevenly-sampled sparse light curves, such as \textit{Gaia}.
\end{abstract}

\keywords{Astronomy data analysis --
		Transit photometry --
		Exoplanet detection methods --
		Algorithms}

\section{Introduction} \label{sec:intro}

The detection of transiting exoplanets is a very active research topic, with implications ranging from physics to astrobiology. Since the first discovery of a transiting exoplanet \citep{Charbonneau1999}, various approaches have been applied to detect the telltale signals of exoplanetary transits in the observed stellar light curves. These approaches mostly focus on the detection of periodic dimmings in the light curves \cite[e.g.][]{Defay2001,Kovacs2002,Renner2008, Hippke2019}. The Box-fitting Least Squares (BLS) algorithm \citep[][hereafter \citetalias{Kovacs2002}]{Kovacs2002} is the most widely-used approach for the detection of periodically transiting exoplanets. It is based on a simplified box-shaped model of a strictly periodic transit, which is characterized by only five parameters:
\begin{itemize}\label{ts_params}
	\setlength\itemsep{0.01em}
	\item{\makebox[2cm]{$P$\hfill} Period}
	\item{\makebox[2cm]{$T_0$\hfill} Epoch of mid-transit}
	\item{\makebox[2cm]{$w$\hfill} Transit duration}
	\item{\makebox[2cm]{$M$\hfill} Out-of-transit mean magnitude}
	\item{\makebox[2cm]{$d$\hfill} Transit depth}
\end{itemize}
This simple approximation of a transit shape requires less parameters than a more accurate one \citep[e.g.][]{Mandel2002,Hippke2019}. Furthermore, the two last parameters, $M$ and $d$, are determined analytically by the data and the temporal parameters $P$, $T_0$ and $w$, and thus the search space is three-dimensional and includes only the first three parameters. Therefore, by the principle of parsimony ('Occam's Razor'), it brings about some statistical advantages \citep[e.g.][]{Jefferys1992}, especially when the data are scarce and their quality does not justify a more detailed model. In addition to the statistical aspects, the simplicity also translates to an algorithmic advantage, by requiring a shorter computation time. In large surveys, with many light curves, using an overly complicated model can result in prohibitively long run times. Thus, in the case of low-cadence surveys such as \textit{Gaia} \citep{Gaia2016}, fitting detailed models to each light curve, for the purpose of detection, is even likely to be counterproductive, in both efficiency and run time. It is therefore necessary to come up with an adaptation of BLS specifically designed for such cases, of sparse light curves.

The popularity of BLS resulted in a variety of implementations and improvements \citep[e.g.][]{Grziwa2012,Ofir2014,Hippke2019}. Many of these implementations are based on the original FORTRAN BLS code, in which the light curve is phase-folded according to each trial period and binned in phase. A double iteration is then performed over the bins to test different beginning (ingress) and end (egress) phases of the transit. Equivalently, some implementations \citep[e.g.][]{CollierCameron2006} scan a different parameter space, replacing the double iteration over phase bins with iteration over discrete grids of phase and duration. Binning, or alternatively discrete phase and duration grids, are useful when dealing with high-cadence photometry with thousands of samples, mainly in terms of computing time. However, they necessarily lose part of the information in the data, such as the exact phases of all the samples. In the case of low-cadence light curves containing only hundreds of samples, in which information is more scarce and computing times are shorter, they might only introduce loss of information, without any significant improvement in running time. 

In this paper we present a new BLS implementation that does not rely on binning nor on phase and duration grids. In Sec.\ \ref{sec:SBLS} we present and detail this novel implementation, and in Sec.\ \ref{sec:issues} we discuss a few implementation issues (such as run-time complexity). Section \ref{sec:performance} discusses the performance of our implementation, and we conclude in Sec.\ \ref{sec:conc} with some final comments.

\section{Sparse BLS} \label{sec:SBLS}

The implementation we introduce here, Sparse-BLS\footnote{A Java code is provided in: \url{https://github.com/aviadpi/SparseBLS}} (SBLS), essentially follows the same original approach described in \citetalias{Kovacs2002}, except the data is left unbinned. 
The essence of the BLS approach is the identification of two brightness levels, in and out of transit, and estimate them in order to compute a goodness-of-fit statistic (e.g. $\chi^2$). When scanning trial configurations of $P$, $T_0$ and $w$, only configuration changes that affect the inclusion of points in the transit affect the goodness of fit, mainly by modifying the estimates of the two levels. A naive grid scanning scheme that ignores the phases of the light-curve samples might thus perform unnecessary computations when the configuration change has no practical effect.  Alternatively, depending on the exact sample times, it might miss other configuration changes that take place between grid points and do have an effect. SBLS, on the other hand, focuses only on configuration changes that affect the inclusion of points in the transit. In fact, for a given period, SBLS effectively scans the space of transit phases and durations continuously, by checking only the configurations in which the goodness-of-fit changes due to inclusion and exclusion of points in transit.

SBLS scans a single grid of trial periods, followed by a double iteration over indices in the phase-folded light curve. However, the indices now refer to the \emph{data points} themselves, instead of the phase bins. This requires, of course, sorting the points by phase for each trial period. Once the light curve is sorted by phase, the double iteration looks for the points in transit. We iterate over all data points for the first in-transit point, with index $i_1$. For the last in-transit point, with index $i_2$, we only need to scan points immediately following $ i_1 $, up to some predetermined maximum duration \citep[e.g.][]{Ofir2014}. We recommend setting a lower limit to the number of points in transit, to reduce the chance of false detections caused by spurious outliers. This of course depends on the total number of samples available.
For example, in our test runs (see Sec.~\ref{sec:performance}), we simulated light curves with a very small number of data points -- $100$. We have set a lower limit of three to the number of points in transits. This ensured a reasonable performance while avoiding a substantial increase in spurious detections.

We repeat here the derivations presented in \citetalias{Kovacs2002}. We denote the original time series indices by $j$, while using $i$ for the indices of the time series sorted by phase ('phase-folded indices'). Given a time series of values $x_j$ and errors $\sigma _j$, we first convert it to a zero-weighted-mean dataset with normalized weights $\left(\tilde{x_j},\tilde{w_j}\right)$:
\begin{align} \label{eq:init1}
    w_j &= \sigma _j ^{-2} & W &= \sum \limits_{j=1}^{N} w_j & \tilde{w_j} &= \dfrac{w_j}{W}\\
    \label{eq:init2}
    \mu &= \sum \limits_{j=1}^{N}\tilde{w_j} x_j  & \tilde{x_j} &= x_j -\mu
\end{align}
For each trial period $P$ we compute the phase values of the folded time series using the modulo operation\footnote{We assume that the modulo operation returns positive values for negative dividends. This is not necessarily the case in all programming languages.}:
\begin{equation} \label{eq:phase}
    \phi_i =  \frac{t_i \mod P}{P}
\end{equation}
For a given set of in-transit phase-folded indices $\left\{i_1,i_1+1,\ldots,i_2\right\}$, we calculate the Signal Residue (\textit{SR}):
\begin{align} \label{eq:srSR}
    s &= \sum \limits_{i=i_1}^{i_2} \tilde{w_i} \tilde{x_i} & r &= \sum \limits_{i=i_1}^{i_2} \tilde{w_i} & \mathrm{\textit{SR}} = \left(\dfrac{s^2}{r(1-r)}\right)^{1/2}
\end{align}

The $s$ and $r$ values can be used to estimate the stellar magnitude $\left( M \right)$ and the transit depth $\left( d \right)$:
\begin{align}
    \hat{M} &= \mu - \frac{s}{1 - r} & \hat{d} &= \frac{s}{r \left( 1 - r \right)}
\end{align}

The maximal \textit{SR} value is saved for each trial period, resulting in a figure-of-merit assigned to each period, i.e.\ a periodogram. 

The calculation of the $s$ and $r$ values in SBLS makes use of memoization, which reduces the computing time significantly. Memoization is a common technique to optimize code execution by caching and retrieving interim values of various variables. SBLS benefits from a memoization method we dubbed "sliding window", in which the values of $s$ and $r$ are sequentially updated only once in each loop iteration, by subtracting or adding the necessary value for every incremental change of either $i_1$ or $i_2$. 

\begin{figure}
	\includegraphics[width=1\linewidth]{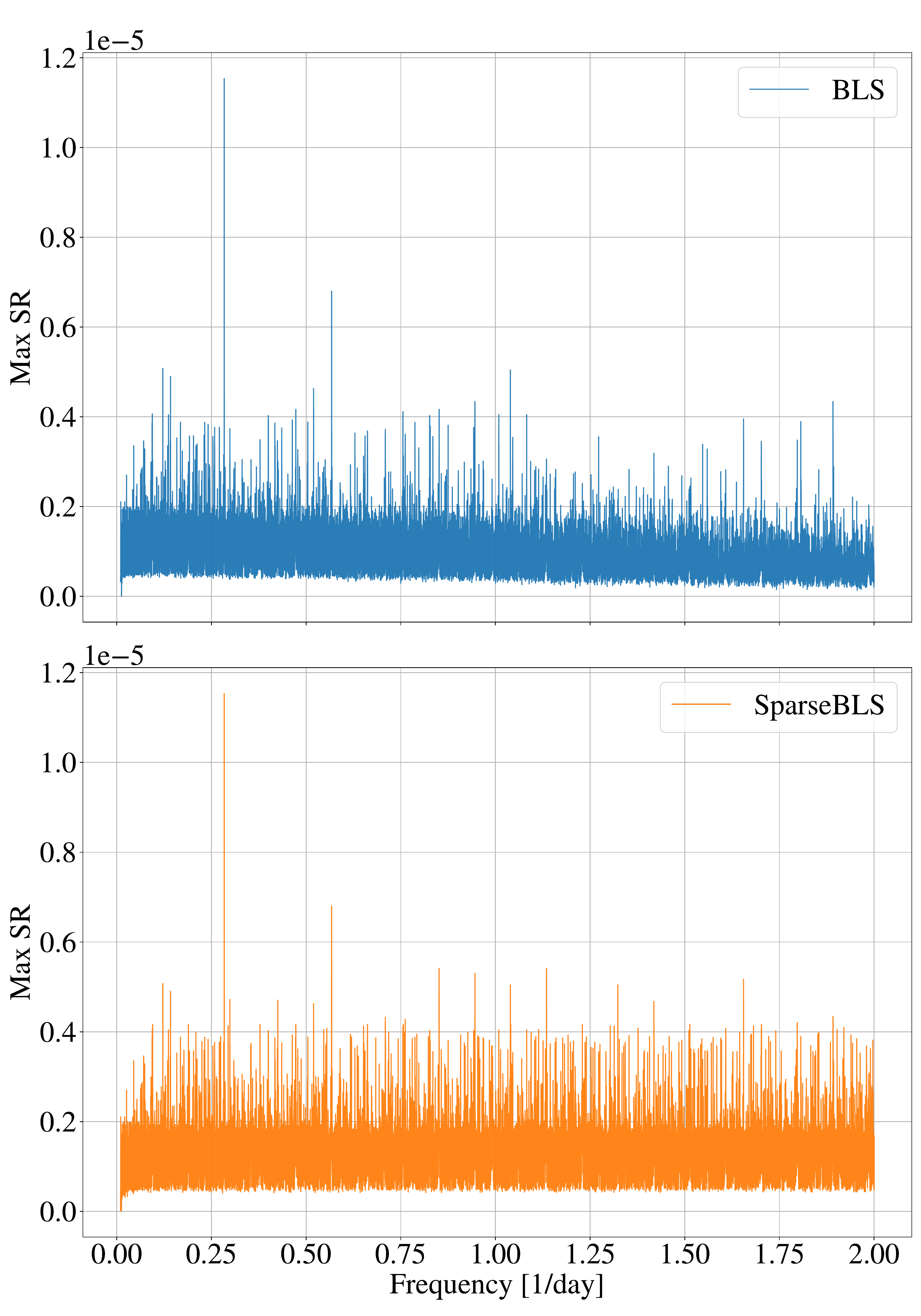}
	\caption{Spectra of BLS (\textit{Top}) and SBLS (\textit{Bottom}). See text for details of the simulated signal. In both spectra the most prominent peak is at the correct frequency.}
	\label{fig:fGrams} 
\end{figure}

Figure~\ref{fig:fGrams} shows the BLS and SBLS spectra obtained for a simulated test case of a low-cadence light curve with $200$ samples over a baseline period of $1\,000$~days and a noise standard deviation of $1$~mmag. The simulated transit parameters are based on those of HD$209458$b , with a period of $3.52472$~days, transit duration of $0.128$~days and a depth of $0.0164$~magnitude \citep{Charbonneau1999}.

Figure \ref{fig:fGrams} demonstrates another feature of SBLS: although the prominent peak in the two spectra is very similar, there is a slight difference manifested as a small slope in the BLS spectrum. This is related to the number of transit configurations tested for each trial period, which is proportional to the period in the specific phase grid scheme we used, and thus fewer configurations are tested in the higher frequencies. The SBLS spectrum does not use any scheme for binning or phase grid, and therefore the number of configurations tested in each period depends only on the number of data points. 

\section{Implementation issues} \label{sec:issues}

\subsection{Run-time complexity} \label{Imp_runtime}
Let us compare the estimated time complexity of the BLS and SBLS implementations. We can safely omit the identical preparatory stages in which we subtract the mean and normalize, especially since their complexity is only $\mathcal{O}(N)$ operations. For the comparison we chose to use the equivalent BLS implementation \citep[e.g.][]{CollierCameron2006}, rather than the implementation that uses binning. This implementation scans a three-dimensional parameter space of periods, phases and durations, with grid sizes of $N_P,N_{T_0},N_w$, respectively. For each configuration BLS then proceeds to calculate \textit{SR}, an $\mathcal{O}(N)$ procedure, resulting in a total time complexity of $ \mathcal{O}(N \cdot N_P \cdot N_{T_0} \cdot N_w)$. Thus, the BLS time complexity depends linearly on $N$ (assuming $N_P,N_{T_0},N_w$ are determined independently of $N$).

SBLS, on the other hand, scans a single grid of trial periods, and for each period orders the samples by their phases, requiring $\mathcal{O}(N\log N)$ operations per trial period. Iteration over the in-transit indices $\left(i_1,i_2\right)$ requires $\mathcal{O}(N^2)$ operations, though this is only an upper limit (and therefore a worst-case estimate), as $i_2$ is usually not too far from $i_1$. The calculation of $SR$ would naively require additional $\mathcal{O}(N)$ steps, but can be reduced to $\mathcal{O}(1)$ using memoization. The total time complexity is therefore $ \mathcal{O}\left(N_P \cdot(N\log N + N^2) \right) \xrightarrow{} \mathcal{O}(N_P \cdot N^2)$. In cases where the phase and duration grid sizes, $N_{T_0}$ and $N_w$, depend on $N$, the difference between BLS and SBLS time complexities would only increase, making BLS less preferable over SBLS for large $N$. However, as we show in Sec.\ \ref{sec:performance}, the balance might shift only at quite large values of $N$, and therefore SBLS still has superior performance for small $N$.

\subsection{Duration of transit}
\begin{figure}
	\includegraphics[width=1\linewidth]{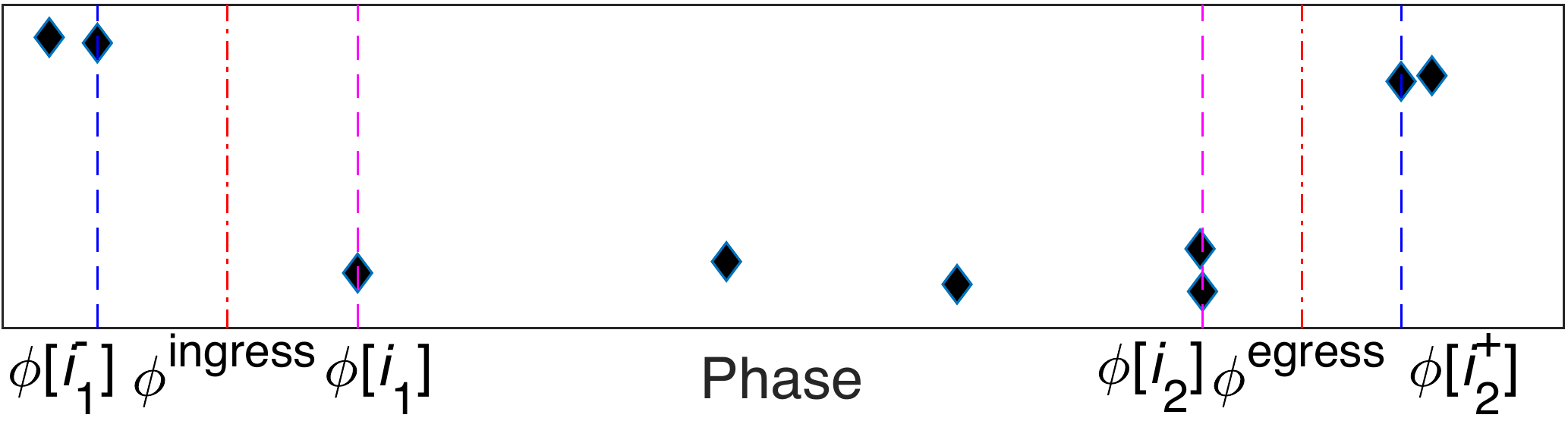}
	\caption{Schematic illustration of the phase calculations in Eqs.\ \ref{eq:i1}--\ref{eq:egress}. The dashed magenta lines mark the phases of the first and last samples in the transit ($i_1$ and $i_2$), while the dashed blue lines mark the phases of the neighbouring samples out of transit. The red dashed-dotted lines mark the estimated phases of the ingress and egress obtained by averaging the phases of the corresponding in-transit and out-of-transit phases.}
	\label{fig:Phases} 
\end{figure}

In this paper, phase values, denoted by $\phi$, are defined so that they are dimensionless numbers ranging between $0$ and $1$. For any pair of transit-scanning indices in the phase-folded light curve, $\left(i_1,i_2\right)$, we find their out-of-transit neighbours $\left(i_1^-,i_2^+\right)$ and use them to estimate ingress and egress phase values, while accounting for wrap-around phases: 
\begin{align} \label{eq:i1}
& \quad \quad \quad \quad \quad i_1^- = \left( i_1 - 1 \right)\mod N\\
& \quad \quad \quad \quad \quad i_2^+ = \left( i_2 + 1 \right) \mod N
\end{align}
\begin{align} \label{eq:ingress}
    \phi^\text{ingress} = 
    	\begin{cases} \frac{\phi[i_1]+\phi[i_1^-] }{2} &\text{if\quad} \phi[i_1^-]<\phi[i_1]\\\\
     \frac{\phi[i_1]+\phi[i_1^-] }{2} - \frac{1}{2} &\text{if\quad} \phi[i_1^-]>\phi[i_1]
        \end{cases}
\end{align}
\begin{align}
\label{eq:egress}
    \phi^\text{egress} = 
        \begin{cases} \frac{\phi[i_2]+\phi[i_2^+] }{2} &\text{if\quad} \phi[i_2]<\phi[i_2^+]\\\\
        \frac{\phi[i_2]+\phi[i_2^+] }{2} +  \frac{1}{2} &\text{if\quad} \phi[i_2]>\phi[i_2^+]
        \end{cases}
\end{align}
Figure \ref{fig:Phases} serves to illustrate schematically the calculations detailed above.

Finally we estimate the transit duration in units of time, using $\phi^\text{ingress}$ and $\phi^\text{egress}$:
\begin{align} \label{eq:duration}
	w &= \left[ \left( \phi^\text{egress}-\phi^\text{ingress}\right) \mod 1 \right] \cdot P
\end{align}

Figure \ref{fig:sbls} provides a pseudocode of the SBLS algorithm, depicting the steps necessary for memoization, using the above definitions of the ingress and egress phases, and the duration.

\begin{figure} 
    \input{PseudoCode}
    \caption{SBLS algorithm}
    \label{fig:sbls}
\end{figure}

\subsection{Nearly sorted time series}
\label{sec:sorting} 

In cases where the frequency grid is very dense, e.g.\ frequency steps of $10^{-5}\,\text{~d}^{-1}$ with a time baseline of $T \sim 1\,000$~d, the phase-folded order of the samples changes only slightly for successive trial frequencies. In such cases, the sorting time complexity can be reduced by applying sorting techniques tailored for nearly-sorted arrays on the light curve which is already partially sorted from the previous trial frequency \citep[e.g.\ insertion sort;][]{CooKim1980}. However, since the conventional sorting time complexity, $\mathcal{O}(N\log N)$, is negligible compared to the time spent in the double iteration, $\mathcal{O}(N^2)$, the improvement in run time is only on the order of a few percent of the total run time (less than $5\%$ in our tests).

\section{Performance} \label{sec:performance}

\subsection{Detection efficiency} \label{sec:detection_eff}
As \citetalias{Kovacs2002} have already shown, binning has the immediate effect of introducing some dependence of the performance on the transit phase, since the transit phase need not necessarily align with the arbitrary binning scheme. This is also the case for algorithms that scan discrete phase and duration grids. By avoiding arbitrary binning or phase grids, the performance of SBLS is by definition independent of the transit phase. 

We present here one of the tests we have performed for comparing the performance of BLS and SBLS, by applying them on two separate datasets -- one containing pure white Gaussian noise without any signal, and another containing randomly generated transit signals at a predetermined signal-to-noise ratio (SNR) as defined by \citetalias{Kovacs2002}. For each dataset we simulated $1\,000$ light curves, each containing $100$ points with uniformly-distributed random sampling times in the range of $\left[ 0,1\,000 \right]$~days. We generated transit signals with a transit depth of $d=0.01$~mag, noise with standard deviation of $\sigma = 3$~mmag, using frequencies drawn from a log-uniform distribution in the range $\left[ 0.1,2.0 \right] \mathrm{~d}^{-1}$, transit phases drawn from a uniform distribution in the range $\left[ 0,1 \right]$ and transit duration set to $3\%$ of the drawn period (shorter duration might not be detectable at all either by BLS or by SBLS with only $100$ samples). Both BLS and SBLS \textbf{scanned} frequencies in the same range as the simulated frequencies, with a frequency step of $10^{-4} \mathrm{~d}^{-1}$.  For the BLS we used a phase grid size of $N_{T_0} = 1\,000$ and $N_w \leq 16$ for the transit duration grid, ranging between $0.02\mathrm{~d}$ and the upper limit for the transit duration, used in both implementations. The upper limit we used for the duration was proportional to $P^{1/3}$, similarly to that used by \cite{CollierCameron2006}. We used the maximum $SR$ value as the detection statistic. 

We chose to ignore the correctness of the frequency where the maximum $SR$ was obtained, since if there were only few points in transit (due to the sparse sampling), there could be several equivalently plausible frequencies that led to the same affiliation of points to transit. In real life this issue should be resolved by follow-up observations. Thus, a true detection was declared when the returned maximum $SR$ value exceeded the threshold value when applied on light curves containing a signal, disregarding its frequency.

\begin{figure}
	\includegraphics[width=0.925\linewidth]{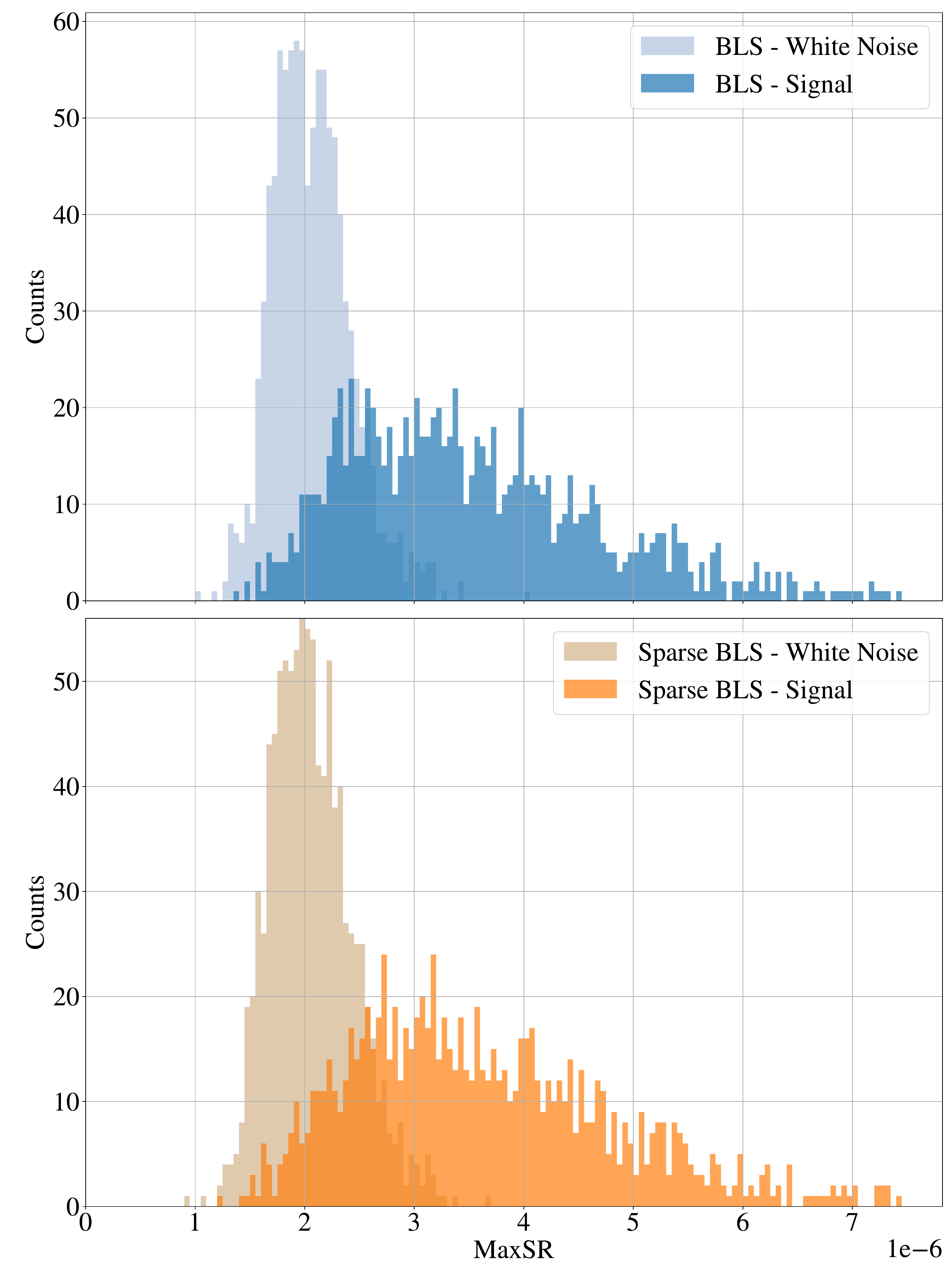}
	\caption{Sample distributions of the Maximum SR values for datasets containing white noise and transit signals, of BLS (\textit{top}) and Sparse-BLS (\textit{bottom}).}
	\label{fig:MaxSR_dists} 
\end{figure}

We ended up with four distributions for the test statistic -- "Noise" and "Signal" for each one of the two implementation. Figure \ref{fig:MaxSR_dists} presents histograms of the four distributions. Apparently, the two implementations produced very similar distributions, as expected. In both of them there is considerable overlap between the Signal and Noise histograms. This is related to the relatively difficult detection challenge -- only $100$ samples and considerable noise. When the number of samples increases, the two distributions are presenting diminishing overlap.

\begin{figure}
	\includegraphics[width=0.95\linewidth]{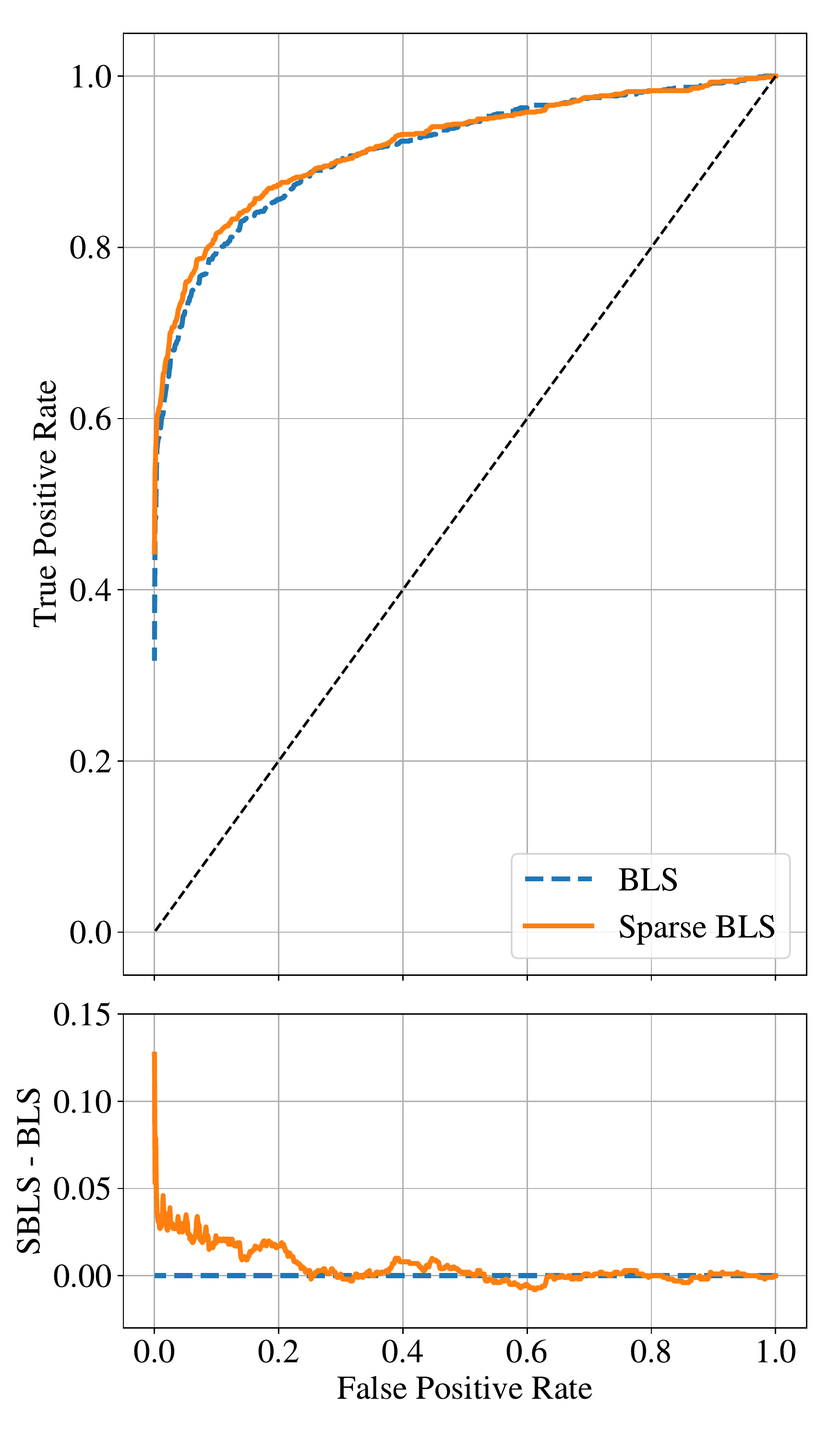}
	\caption{\textit{Top:} Comparison of the ROC curves of BLS and SBLS. \textit{Bottom:} The difference between the two ROC curves. The details of the simulations used in producing the curves are listed in the main text.}
	\label{fig:ROC} 
\end{figure}

The Noise and Signal distributions described above can be used to compile a Receiver Operating Characteristic (ROC) curve. A ROC curve \cite[e.g.][]{FAWCETT2006861} is a graphical tool to illustrate the performance of a detection scheme, and is widely used in signal detection theory and related disciplines. As the detection threshold increases, the rate of positive detection decreases -- both true (correctly identifying a transit signal as such) and false (wrongly identifying a pure noise light curve as one that contains transit). The ROC curve presents the way the true positive rate and the false positive rate are related, by scanning different detection thresholds and counting the number of true and false detections above them.
Figure \ref{fig:ROC} presents the two ROC curves resulting from the distributions in Fig.~\ref{fig:MaxSR_dists}. The upper panel presents the two ROC curves, while the lower panel shows the difference between them. It is clearly seen that the SBLS curve is dominating the BLS curve, i.e.\ the SBLS true positive rate is larger than that of the BLS for almost every assumed false positive rate. This is a manifestation of the non-sensitivity of the SBLS performance to the transit phase. Of course, using a much finer grid for the transit phase and duration may eventually improve the performance of BLS, but that would come on the expense of running time as was shown above in Sec.~\ref{Imp_runtime}.

\subsection{Run time comparison}
\begin{figure}
\includegraphics[width=1.\linewidth]{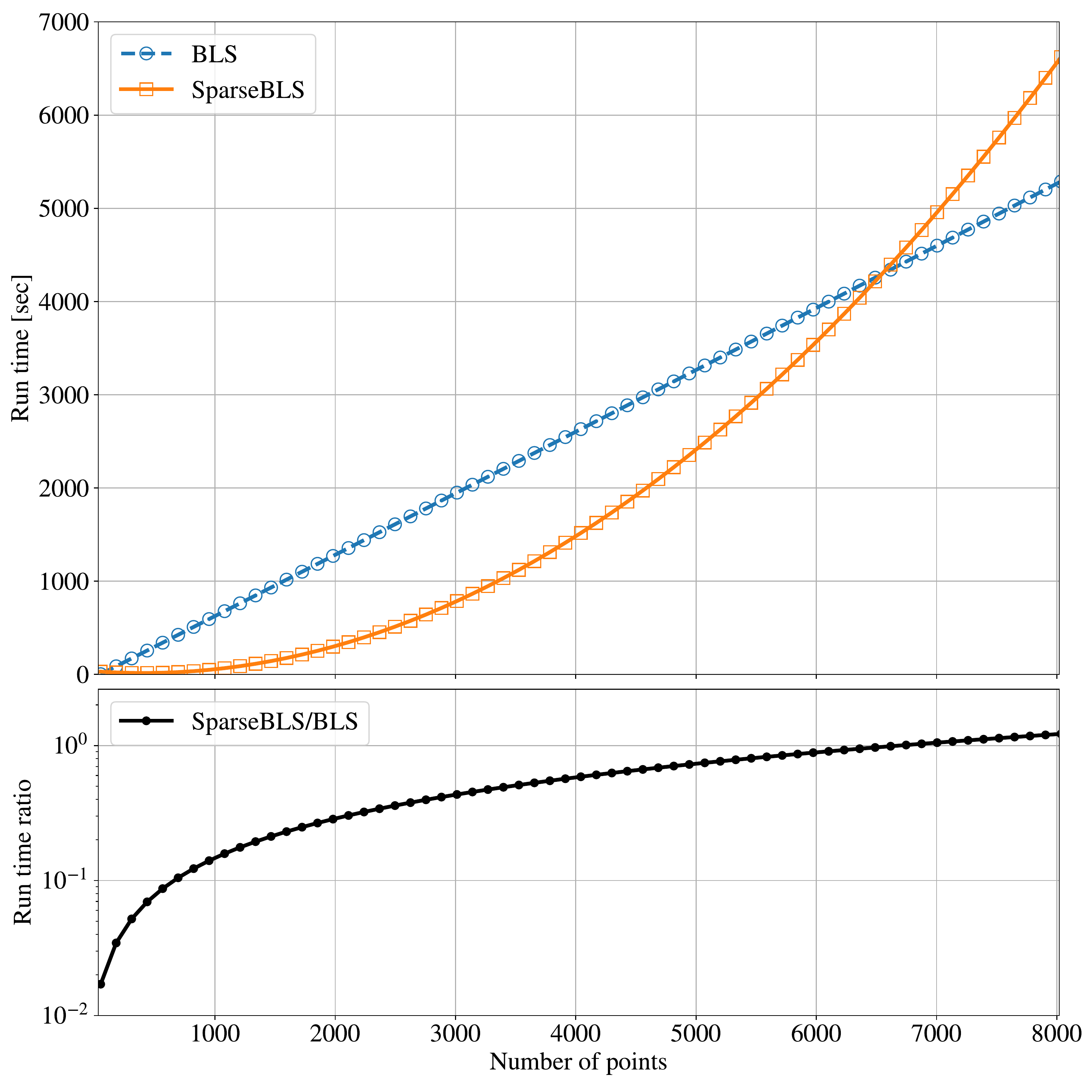}
\caption{\textit{Top:} Run time of BLS and SBLS for light curves with up to $8\,000$ points. \textit{Bottom:} Run time ratio: SBLS over BLS.}
\label{fig:Runtime_benchmark}
\end{figure}

We have implemented both BLS and SBLS in Java, and measured the run times of both implementations for different sizes of datasets, using a $3.4$ GHz Intel Octa-core i7 CPU. Figure~\ref{fig:Runtime_benchmark} presents the results of this comparison. As expected, one can see a linear dependence of the BLS run time on the size of the light curve, and a quadratic dependence for SBLS. Indeed, for large light curves SBLS is slower than BLS. However for sparse light curves, with less than $5\,000$ points, the SBLS run time can be considerably shorter than that of the BLS, even down to $2\%$.

\section{Conclusions} 
\label{sec:conc}

We have presented the SBLS (Sparse BLS) implementation of the BLS. Its main advantage is that it is independent of any arbitrary phase grid or binning scheme, unlike the original BLS of \citetalias{Kovacs2002}. However, it is not a very efficient implementation in terms of computing time, and therefore it might not be suitable to long, high-cadence surveys, such as Kepler \citep{Boretal2010} or TESS \citep{Ricetal2014}.

Nevertheless, the dependence of the complexity on the size of the light curve is quadratic, and curiously enough, for data that is sparse enough, of less than a few thousand samples per light curve, the linear dependence of the BLS complexity is still considerably larger than the quadratic complexity for SBLS. In those cases, the resource balance tilts towards SBLS, and it becomes clear that SBLS is more efficient than BLS, both in terms of complexity and in terms of detection efficiency. 
In fact, our complexity estimates of the BLS runtime, which we presented in Sec.~\ref{Imp_runtime}, were a little simplistic in assuming that the number of bins (or alternatively grid cells) does not depend on the light-curve size ($N$). If we introduce a dependence of $N_{T_0}$ and $N_w$ on $N$, this would make the $N$-dependence of the BLS complexity superlinear, increasing even more the breakeven point in the upper panel of Fig.~\ref{fig:Runtime_benchmark}.

Thus, for large and sparse datasets such as \textit{Gaia}, SBLS should be the tool of choice for searching for periodic planetary transits.\\

\acknowledgments
This research was supported by the ISRAEL SCIENCE FOUNDATION (grant No. 848/16). We also acknowledge partial support by the Ministry of Science, Technology and Space, Israel. 

\bibliography{SBLS}{}
\bibliographystyle{aasjournal}

\end{document}

%% file: PseudoCode.tex
\SetAlCapSkip{-1em}
\begin{algorithm}[H]
	\KwIn{Light curve \textit{LC}$=\{(t_j, x_j, \sigma_j) \, | \,j=0 \ldots N-1 \}$}
	\KwOut{\textit{\textbf{Periodogram}}, \textit{\textbf{in-transit indices}}}
	Initialize \textit{LC};\\
	\For{\upshape trial period $P$} {
		Sort \textit{LC} by phase with $P$ (Eq.\ \ref{eq:phase}): $\left( \phi_i,x_i\right)  \gets \left( t_j, x_j \right)$  \\
		$ {s},{r},{s}_\text{prev},{r}_\text{prev} \gets 0 $\\
		\For {\upshape{phase index} $ i_1 \in  [0,N-1] $ }
		{
			$ {w} \gets {0}$\\
			$ {s} \gets {s}_\text{prev}$\\
			${r} \gets {r}_\text{prev}$\\
			$ i_2 \gets i_{2,\text{prev}} $\\
			\If{$ i_1 > 0 $}{
			    subtract values in $i_1^-$ and $i_2 $ from $ {s,r} $ \\
			}
			estimate $\phi^\text{ingress}$ using Eq.\ \ref{eq:ingress}\\
			\While{$w  < $ \textit{maxDuration}}{
				\If{\upshape too few points in transit}{
				    add values in $i_2$ to $ {s,r} $\\
					$ {s}_\text{prev} \gets {s}$\\
					${r}_\text{prev} \gets {r}$\\
					$ i_{2,\text{prev}} \gets i_2 $\\
					$ i_2 \gets (i_2+1) \mod N  $\\
					continue
				}
				add values in $i_2$ to $ {s,r} $\\
				estimate $\phi^\text{egress}$ using Eq.\ \ref{eq:egress}\\
				estimate $w$ using Eq.\ \ref{eq:duration}\\
				calculate \textit{SR} using $ {s,r} $\\
				\If{\textit{SR} $>$ \textit{MaxSR}}
				{
					\textit{MaxSR} $\gets$ \textit{SR} \\
					save \textit{\textbf{in-transit indices}}\\
				}
				$ i_2 \gets (i_2+1) \mod N $
			}
		}
        	save \textit{MaxSR} to \textit{\textbf{Periodogram}}\\
        	\textit{MaxSR} $\gets 0 $
        	}
        	\Return{\textbf{\textit{Periodogram}}, \textbf{\textit{in-transit indices}}}
\end{algorithm}
        